\documentclass[fleqn,10pt]{wlscirep}
\usepackage[english]{babel}
\usepackage[babel]{csquotes}
\usepackage{caption}
\usepackage{soul,xcolor}
\usepackage[symbol]{footmisc}
\setstcolor{red}

\newcommand{\red}[1]{\textcolor{red}{#1}}

\newcommand{\ovl}[1]{\overline{#1}}
\newcommand{\lag}{\mathrm{lag}}
\newcommand{\beq}{\begin{equation}}
\newcommand{\eeq}{\end{equation}}

\newcommand{\CV}{\mathrm{CV}}
\newcommand{\tlag}{t_{\mathrm{lag}}}
\newcommand{\tlog}{t_{\mathrm{log}}}
\newcommand{\lmax}{\lambda_{\max}}
\newcommand{\Flag}{\mathcal{F}_{\mathrm{lag}}}

\title{Initial cell density encodes proliferative potential in cancer cell populations}

\author[1,2]{Chiara Enrico Bena}
\author[2,3]{Marco Del Giudice}
\author[4,9]{Alice Grob}
\author[2]{Thomas Gueudr\'e}
\author[5]{Mattia Miotto}
\author[6]{Dimitra Gialama}
\author[7]{Matteo Osella}
\author[8]{Emilia Turco}
\author[6,9]{Francesca Ceroni}
\author[2,10]{Andrea De Martino}
\author[11,2,*]{Carla Bosia}

\affil[1]{Sorbonne Universit\'e, CNRS, Institut de Biologie Paris-Seine (IBPS), Laboratoire Jean Perrin (LJP), F-75005, Paris, France}
\affil[2]{IIGM - Italian Institute for Genomic Medicine, c/o IRCCS, Str. Prov.le 142, km 3.95, 10060, Candiolo, Italy}
\affil[3]{Candiolo Cancer Institute, FPO - IRCCS, Str. Prov.le 142, km 3.95, 10060, Candiolo, Italy}
\affil[4]{Department of Life Sciences, Imperial College London, London, UK}
\affil[5]{Department of Physics, Sapienza University, Piazzale Aldo Moro 5, 00185, Rome, Italy}
\affil[6]{Department of Chemical Engineering, Imperial College London, London, UK}
\affil[7]{Physics Department and INFN, University of Turin, Via P. Giuria 1, 10125, Turin, Italy}
\affil[8]{Molecular Biotechnology Center, University of Turin, Via Nizza 52, 10126, Turin, Italy}
\affil[9]{Imperial College Centre for Synthetic Biology, London, UK}
\affil[10]{Soft \& Living Matter Lab, CNR-NANOTEC, Rome, Italy}
\affil[11]{Department of Applied Science and Technology, Politecnico di Torino, Corso Duca degli Abruzzi 24, 10129, Turin, Italy}
\affil[*]{Corresponding author: carla.bosia@polito.it}


\begin{abstract}
Individual cells exhibit specific proliferative responses to changes in microenvironmental conditions. Whether such potential is constrained by the cell density throughout the growth process is however unclear. Here, we identify a theoretical framework that captures how the information encoded in the initial density of cancer cell populations impacts their growth profile. By following the growth of hundreds of populations of cancer cells, we found that the time they need to adapt to the environment decreases as the initial cell density increases. Moreover, the population growth rate shows a maximum at intermediate initial densities. With the support of a mathematical model, we show that the observed interdependence of adaptation time and growth rate is significantly at odds both with standard logistic growth models and with the Monod-like function that governs the dependence of the growth rate on nutrient levels. Our results (i) uncover and quantify a previously unnoticed heterogeneity in the growth dynamics of cancer cell populations; (ii) unveil how population growth may be affected by single-cell adaptation times; (iii) contribute to our understanding of the clinically-observed dependence of the primary and metastatic tumor take rates on the initial density of implanted cancer cells. 
\end{abstract}

\begin{document}
\maketitle

\section*{Introduction}

The so-called ``lag'' phase that precedes the exponential growth of a cell population  entails the adaptation of cells to the growth medium. Its duration (the ``lag time'') is known to be affected by a number of factors, including the history of the population prior to inoculation \cite{baranyi1994dynamic,delignette1998relation,dufrenne1997effect}. Initial conditions, including pre-inoculation ones, are however thought to be gradually forgotten as cells adapt, so that the growth rate characterizing the subsequent exponential (or ``log'') regime only encodes the current physiological state of cells and environmental conditions. In this respect, the growth rate of a cell population provides the most elementary measure of its fitness. It might then be somewhat surprising that both the lag phase and the growth rate can be affected rather significantly by the density $N_0$ of the initial population (the {\it inoculum}). 

Following Rein \& Rubin's pioneering work \cite{Rein1968}, inoculum-density dependent traits have been observed, among others, in bacterial \cite{Postma1990,Coleman2003,Irwin2010,Koutsoumanis2013}, insect \cite{Marteijn2008}, plant,  \cite{Gulik1994,Kanokwaree1997,Carvalho1999} and cancer cells \cite{Gregorio2016,N.Rodriguez2001,Ozturk1990}, most notably impacting metabolism (specifically the ability to excrete certain compounds \cite{Ozturk1990}), carrying capacity  \cite{Ozturk1990}, and the lag time \cite{Pin2006}. For instance, $N_0$-dependent lag times have been reported at very low inoculum densities \cite{Augustin2000} or under stress, when only small sub-populations can sustain growth \cite{Robinson2001}. In turn, the growth rate was found to increase \cite{Marteijn2008,Carvalho1999} or decrease \cite{Kanokwaree1997,Gregorio2016} with the initial density depending on the organism and the growth medium. Inoculum-density dependence has also been implicated in the growth rate's sample-to-sample variability \cite{Irwin2010,Gulik1994} and, more recently, in the overall growth kinetics of a population of cancer cells\cite{johnson2019cancer}.

While intriguing, $N_0$-dependencies are to be expected, at least from a theoretical viewpoint. For instance, under logistic growth, the maximum growth rate of a population of non-interacting cells naturally declines as $N_0$ approaches the carrying capacity of the medium  \cite{baranyi1994dynamic,DeMartino2017}. Other types of behaviours, especially positive correlations between the growth rate and $N_0$, may have different origins. For instance, it is known that randomness in cellular reproduction events can propagate to macroscopic parameters \cite{Taheri-Araghi2015a,Jafarpour2018,Hashimoto2016}. If such heterogeneities are putatively averaged out in large inocula, they may become relevant when the initial cell density is small and/or when the performance of the population is driven by single cells having extreme behaviours. Purely stochastic effects indeed explain why identically prepared single cells can generate very different growth trajectories \cite{Koutsoumanis2013}. Likewise, higher fitness could be achieved by populations starting from larger inocula through strengthened cooperation\cite{Stephens1999,johnson2019cancer}. How these features, putatively established during the adaptation phase, are carried into the exponential phase is however not always clear. 

One of the factors that limit our understanding of inoculum-density dependencies lies in the lack of experimental studies characterizing the growth of a cell population with quantitative accuracy across a broad range of initial densities. In this work we provide such a study. In specific, we analyzed the complete growth dynamics (from inoculum to saturation) of a large number of populations of two widely used cancer cell lines, namely Jurkat and K562, growing in fixed carbon-limited media from initial densities ranging over 5 orders of magnitude. We found that the average growth rates of the populations has a striking non-monotonic dependence on $N_0$, with a plateau at small $N_0$, the theoretically expected logistic decrease at large $N_0$, and a marked maximum at intermediate values. Such a behaviour contrasts markedly with the (monotonic) Monod-like function that governs the dependence of the growth rate on nutrient levels \cite{MacIver2008}. Growth rate fluctuations, on the other hand, get larger as the inoculum density increases. The mean lag time and lag time fluctuations were instead found to decrease as the initial cell density increases. For a quantitative insight, we calibrated a simple mathematical model with cues from the growth-curve fitting function to derive relationships linking macroscopic growth parameters to each other through $N_0$. Our theory suggests that growth properties are importantly affected by heterogeneities in single-cell adaptation times. In addition, we propose that the joint effect of a finite carrying capacity and weak cooperative effects during growth can explain the non-trivial behaviour of the maximum growth rate. Implications of our results, both for cancer biology and for theoretical approaches, are discussed in the final part of this article. 

\section*{Results}

\subsection*{Automated quantitative analysis of Jurkat growth curves}

We performed batch culture experiments on a widely studied cancer cell line (Jurkat) growing in suspension. Experiments proceeded as follows (see Figure 1a). (i) A sample of density $N_{0}$ cells/ml was taken from a population growing exponentially in a standard medium whose glucose concentration was at saturation of the Monod function \cite{MacIver2008}. (ii) Cells were then seeded to a new dish supplied with a fixed amount of fresh medium of the same quality. (iii) Finally, cells were grown on the fresh medium. Following adaptation (``lag phase''), every population entered a regime of exponential growth (``log phase'') with a roughly constant rate up to the maximum attainable (the carrying capacity of the medium, $k$), at which point the concentration of cells saturated. We monitored the growth dynamics daily until saturation, recording the corresponding growth curve, namely the logarithm of the concentration of viable cells (in units of $N_0$) versus time (see Figure 1b). A total of $217$ populations (i.e. growth curves) starting from different values of $N_0$ was collected. Ultimately, the initial densities we considered span five orders of magnitude, from $N_0\simeq 10^2$ cells/ml to $N_0\simeq 7\times 10^6$ cells/ml. Values of $N_0$ were chosen (within experimental accuracy) so as to span a range as broad as possible in a roughly uniform way. Growth curve parameters, representing respectively the lag time $\tlag$, the maximum growth rate $\lmax$ and the carrying capacity $k$ through the quantity $A\equiv\ln(k/N_0)$, where then inferred from the growth curves (see Figure 1a and 1b and Materials and Methods, as well as Figure S1 in Supplementary Information for the behaviour of $A$ versus $N_0$). The estimated carrying capacity $k$ turned out to have a relatively broad distribution over our ensemble of populations (see Figure 1c), with mean $\ovl{k}\simeq 8.4\times 10^6$. The largest inoculum densities we considered lie just below the mean carrying capacity. 
Figures 1d-f show that the various parameters characterizing growth display weak linear correlations among each other across our experiments.

\begin{figure}[h!]
\begin{center}
\includegraphics[scale=0.8]{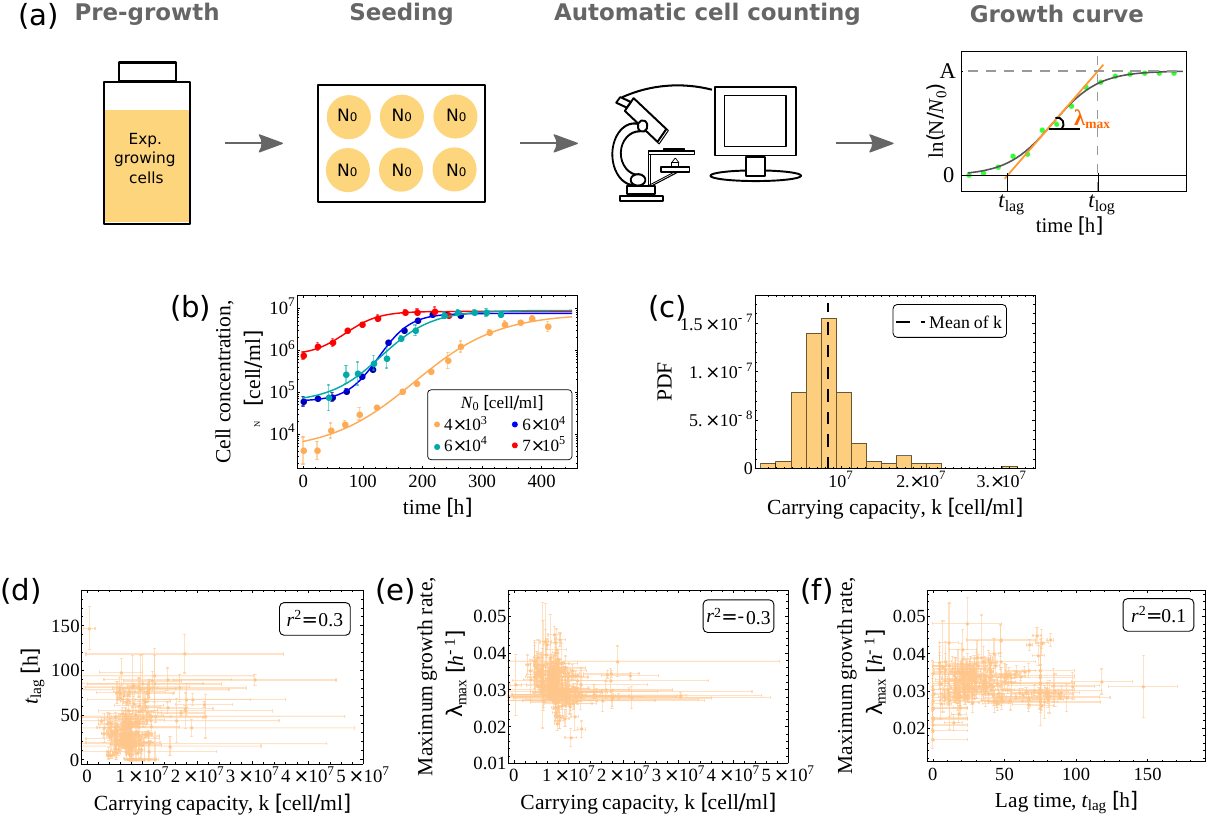}
\end{center}
\caption{(a) Sketch of the experimental procedure employed for Jurkat cells. The right-most panel shows a representative growth curve (green circles) with the corresponding fit (gray sigmoidal curve). The fitting parameters (saturation level $A$, maximum growth rate $\lmax$ and lag time $\tlag$) are displayed. (b) Examples of growth curves for different inoculum densities $N_0$ together with the corresponding fitting curves. Error bars represent the standard errors of the mean counts. (c) Empirical distribution of the carrying capacity $k$ with its mean ($\bar k \simeq 8.4 \times 10^6$) emphasized by the dashed vertical line. (d) Scatter chart showing the estimated lag time $\tlag$ and carrying capacity $k$ for each experiment with their respective standard errors. The two quantities display a weak linear correlation ($r^2 = 0.3$, $p< 0.05$). (e) Scatter chart showing the estimated maximum growth rate $\lambda_{max}$ and carrying capacity $k$ with their respective standard errors. The two quantities display a weak linear correlation ($r^2 = -0.3$, $p< 0.05$). (f) Scatter chart showing the estimated maximum growth rate $\lambda_{max}$ and lag time $t_{lag}$ with their respective standard errors. The two quantities display a weak linear correlation ($r^2 = 0.1$, $p=0.14$).}
\end{figure}


\subsection*{The mean lag time decreases as the initial population size increases}

When plotting the mean $t_{\lag}$ (averaged over experiments) versus $N_0$ we observed a clear decreasing linear dependence on $\ln N_0$ (see Figure 2a). Such a decrease is accompanied by large sample-to-sample fluctuations across the entire range of values of $N_0$. To quantify this variability, we resorted to  the empirical coefficient of variation (CV), namely the ratio between the estimated standard deviation and the estimated mean, of the lag time. This quantity appears to be roughly constant across 4 orders of magnitude in $N_0$ (see Figure 2b), with the final increase reflecting the fact that the mean lag time becomes smaller and smaller as $N_0$ approaches the mean carrying capacity. Recalling that the inoculum was seeded in the same medium in which cells were pre-grown, the existence of a significant lag time suggests some degree of conditioning of the pre-growth medium. Moreover, while considerably more marked, the observed trend of the mean $\tlag$ qualitatively matches that found for low inocula in \cite{Pin2006}, where it was traced back to stochastic effects in the single-cell growth dynamics. We shall explore these points more in depth in the following.

\begin{figure}[h!]
\begin{center}
\includegraphics[scale=0.8]{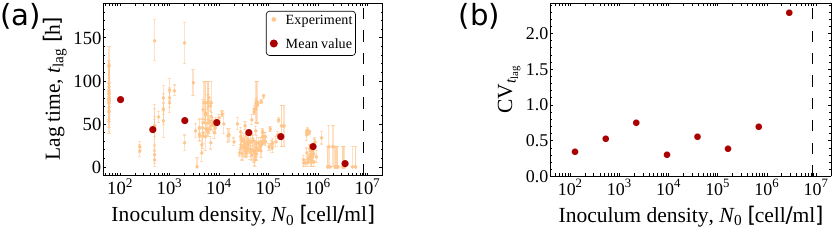}
\end{center}
\caption{(a) Lag time $t_{\lag}$ as a function of the inoculum density $N_0$. Orange markers represent results from individual experiments together with their standard errors. Red markers denote the behaviour of the mean lag time. The dashed black line corresponds to the empirical mean carrying capacity. (b) Empirical behaviour of the coefficient of variation of the lag time ($\CV_{\tlag}$) as a function of the initial density $N_0$.}  
\end{figure}


\subsection*{The mean growth rate is maximized at intermediate initial population size}

The maximum growth rate $\lambda_{\max}$ of a population corresponds to the maximum slope of the growth curve (see Figure 3a). Our experiments returned a broad distribution of values of $\lambda_{\max}$ with a well defined mean but significant variability (up to about 50\%, see Figure 3b). Surprisingly, the mean $\lambda_{\max}$ appears to be modulated by $N_0$ in a complex non-monotonic manner (Figure 3c). For small enough $N_0$ (i.e. $N_0\ll k$), $\lambda_{\max}$ is roughly constant. This agrees with the  growth scenario underpinned by the logistic model as well as by the (weakly cooperative) Allee model, in which initial densities far from the carrying capacity generate populations growing at the rate corresponding to a medium with infinite carrying capacity (the `asymptotic growth rate', see Materials and Methods). Sample-to-sample variability in this regime is generically small, i.e. different populations grow at similar rates, except for the smallest values of $N_0$ ($N_0 \simeq 60$ cells/ml), whose apparently higher dispersion reflects the fact that these initial densities were estimated by subsequent dilutions rather than by direct counting (see Materials and Methods). As $N_0$ increased, we observed a significant enhancement of fluctuations, which noticeably only occur above the reference level given by the asymptotic growth rate. This leads to an overall upward trend in the mean $\lmax$: larger initial densities appear to bear a fitness advantage. Finally, as $N_0$ approaches the carrying capacity, $\lambda_{\max}$ decreases approximately linearly with $N_{0}$, in agreement with the logistic and Allee models (see Figure 3c and Materials and Methods). 

To ensure that these results were not due to our choice of characterizing the exponential phase using $\lambda_{\max}$, we verified that the same qualitative behaviour occurs when the growth rate is read off the growth curves by a linear fit through the  exponential phase (see Figures 3d-f). By construction, such an estimate of the growth rate cannot exceed $\lambda_{\max}$. Results are otherwise qualitatively identical. Likewise, we verified that the observed behaviour of $\lambda_{\max}$ is not due to fluctuations in the carrying capacity across experiments. No significant interdependence between $\lambda_{\max}$ and $k$ was observed (see Figure 1e).

\begin{figure}
\begin{center}
\includegraphics[scale=0.8]{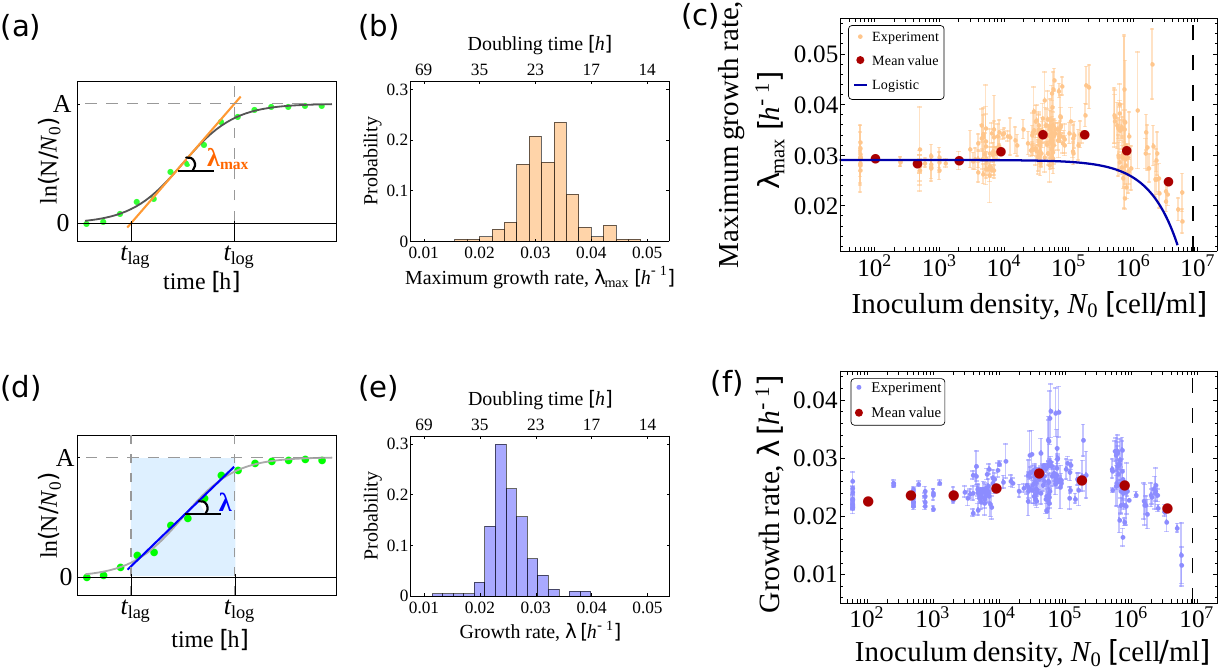}
\end{center}
\caption{(a) Representative growth curve. The fitting parameters are $\tlag$, $A$ and $\lmax$. The maximum growth rate $\lmax$ corresponds to the slope of the tangent to the inflection point (orange line) of the fitting curve (gray curve). The intersection of such a tangent with the time-axis gives the lag time $\tlag$, while its intersection with the line $\ln(N/N_0)=A$ yields the time of exit from the exponential phase, $\tlog$. (b) Empirical distribution of the maximum growth rate. (c) Maximum growth rate $\lmax$ as a function of the inoculum density $N_0$. Orange dots represent parameter estimates from individual experiments with their standard errors, while red dots represent the mean values of $\lmax$. The dashed vertical line marks the value of the mean carrying capacity. The blue line denotes the behaviour of $\lmax$ vs $N_0$ expected on the basis of a purely logistic model with $k = 8.4 \times 10^6$ and asymptotic growth rate $r = 0.029$ (see Materials and Methods). (d) Representative fit of a growth curve using an alternative measure of the growth rate $\lambda$, given by the slope of the linear fit of the data points within the exponential phase of growth (light blue region). (e) Empirical distribution of the alternative growth rate defined in panel (d). (f) Estimated alternative growth rate versus $N_0$. Light blue dots represent results from individual experiments with their standard errors, dark red dots represent their mean values. The dashed vertical line marks the value of the mean carrying capacity.}
\end{figure}


\subsection*{The growth dynamics of K562 cells recapitulates the scenario obtained for Jurkat cells} 

To test whether the scenario just discussed is specific of Jurkat cells, we performed the same study on a different suspension cancer cell line, namely K562 (see Figure 4a for a sketch of the experimental procedure and Materials and Methods for details). We obtained $83$ growth curves for values of $N_0$ spanning more than three orders of magnitude. Growth dynamics was again characterized by fitting growth curves to a sigmoidal function with free parameters related to the carrying capacity ($k$, see Figure 4b-d), growth rate ($\lambda$, Figure 4e) and lag time ($t_{\lag}$, Figure 4f). Despite quantitative differences, the qualitative scenario we obtained is in very good agreement with that found for Jurkat cells. In particular, (i) $\lambda$ displays a clear maximum as a function of $N_0$, with increased fluctuations around it, and (ii) the lag time shows a rough decreasing trend when $N_0$ is increased. Because of a limited carrying capacity, we were unable to resolve the low-$N_0$ plateau of $\lambda$ that characterized the growth rate profile of Jurkat cells. Unfortunately, the values of $N_0$ at which such a plateau should be found ($N_0\lesssim 100$ cells/ml) for K562 cells are inaccessible with our experimental protocol.

\begin{figure}
\begin{center}
\includegraphics[scale=0.8]{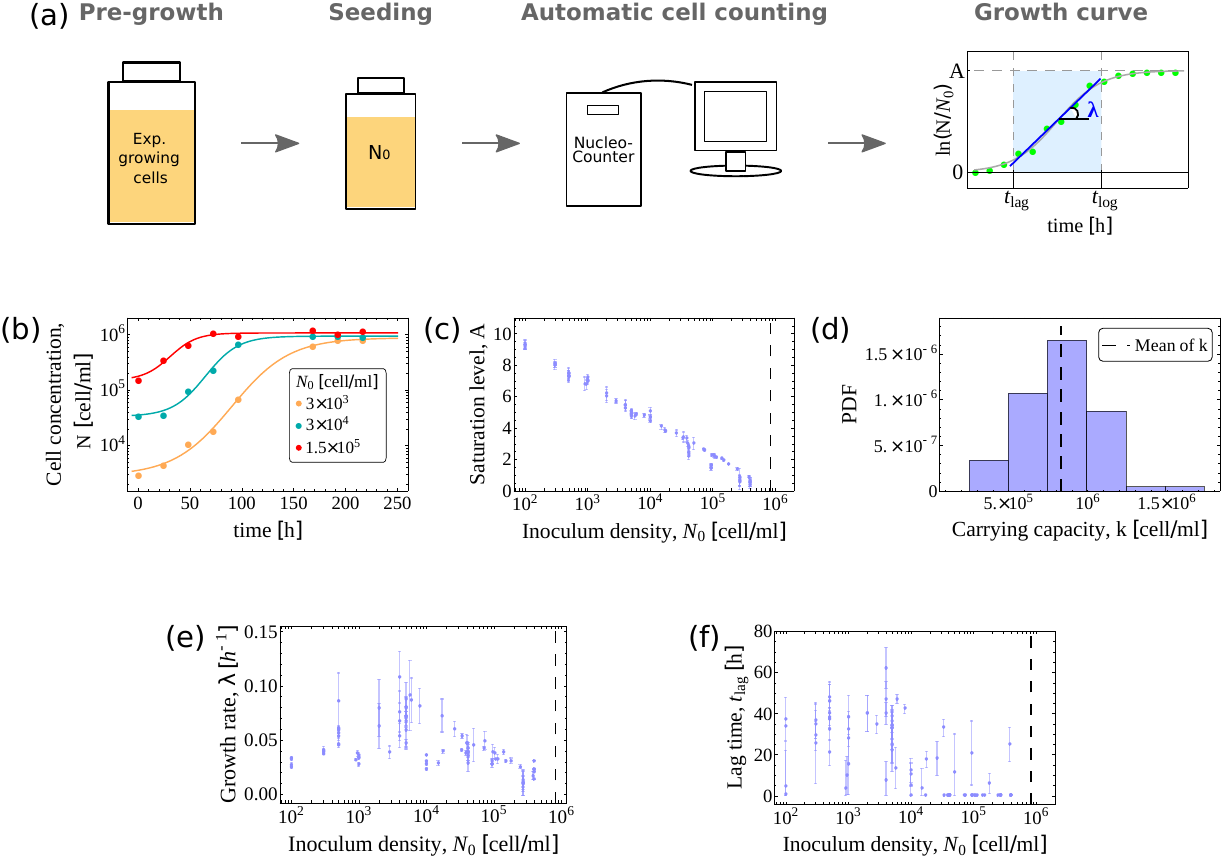}
\end{center}
\caption{(a) Sketch of the experimental procedure employed for K562 cells. The panel on the right shows a representative growth curve (green circles) with the corresponding fit (gray sigmoidal curve). The fitted parameters (growth saturation $A$, growth rate $\lambda$ and lag time $t_{\lag}$) are emphasized explicitly. (b) Three representative growth curves with different inoculum sizes $N_0$ with their corresponding fits. Data error bars are given by $\sqrt{N}$, where $N$ is the concentration of cells. Their size is smaller than the dots, thus they are not visible. (c) The growth saturation $A$, related to the carrying capacity $k$ through the relation $A\equiv\ln(k/N_0)$, is plotted against the inoculum size $N_0$. The mean value of the carrying capacity $k$ is shown by the vertical dashed line. (d) Empirical distribution of the carrying capacity $k$. The dashed vertical line represents the mean of the distribution. (e) Growth rate $\lambda$ as a function of the inoculum size $N_0$. Each light blue dot is an estimated $\lambda$ with its standard error. The dashed black line corresponds to the mean carrying capacity. (f) Adaptation time $t_{\lag}$ as a function of the inoculum size $N_0$. Light blue dots represent the estimated lag times together with their standard errors. The dashed black line corresponds to the mean carrying capacity.}
\end{figure}


\subsection*{Quantitative relationships between growth parameters}

To get quantitative insight into these results, we obtained expressions linking the macroscopic parameters of growth curves (most notably $\tlag$ and $\lmax$) through the inoculum density $N_0$ by constraining a minimal mathematical model of an exponentially growing population with cues from the growth-curve fitting function. Our model assumes that cells in the inoculum are divided in two subpopulations: fast adapters, who begin to expand right after inoculation, and slow adapters that remain quiescent after inoculation for times  longer than the lag time of the culture (see Materials and Methods). Under this simplifying scenario, we found that $\lmax$ and $\tlag$ are approximately related by 

\begin{equation}\label{ltlag}
\lmax\tlag\simeq\ln\left[\left(\frac{k}{N_0}\right)^\alpha-1+p\right]-\ln p~~,
\end{equation}  
where $\alpha\equiv(1+e^2)^{-1}\simeq 0.12$ and $p$ denotes the fraction of fast adapters. Likewise, we found the ``log time'' $\tlog$ at which the population exits the exponential phase to be related to $\lmax$ by
\begin{equation}\label{ltlog}
\lmax \tlog\simeq\ln\left[\left(\frac{k}{N_0}\right)^\alpha-1+p\right]+\ln\frac{k}{pN_0}~~,
\end{equation} 
In spite of the crudeness of our model, these expressions capture empirical results remarkably well with the single adjustable parameter $p$ taking the value $p\simeq 0.4$ (Figure 5a-b). In addition, they allow to estimate the behaviour of the mean lag time and maximal growth rate (Figure 5c-d), along with the respective  coefficients of variation, as functions of $N_0$ (see Figure 5e, Materials and Methods and Supplementary Information, Figures S9 and S10) This suggests that  heterogeneities in single-cell adaptation times are crucial to explain the interrelations between macroscopic growth parameters uncovered by our experiments, a scenario close to that considered in \cite{Pin2006}. However, the parameter $p$ by itself does not appear to fully capture their empirical variability. More refined models e.g. with additional sources of variability and structured (as opposed to bipartite) initial populations are likely to perform even better, albeit at the cost of introducing more adjustable parameters. To conclude we note that, upon conditioning on $N_0$,  our data suggest a peculiar relationship between the mean lag time and the standard deviation of the maximal growth rate estimated across different populations (see Figure 5f): the latter tend to increase as the former gets smaller. In other words, there exists a trade-off between the typical relaxation times and the magnitude of growth rate fluctuations in an ensemble of cell populations growing in homogeneous media. Contrary to similar trade-offs that have been discussed in the context of individual populations\cite{kotte2014phenotypic,chu2016lag,basan2020universal,heinemann2020implications} (i.e. in terms of the variability of growth rates across cells within a given population), trade-offs arising in ensembles of populations represent purely statistical (`emergent') laws whose origin is unclear. Most likely, such laws result from the coupling between the natural variability that characterizes individual cell populations and exogenous stochastic factors (e.g. environmental fluctuations). In this respect, recently discussed stochastic models of population growth that generate robust scaling relationships may yield useful hints\cite{de2016growth,de2016asymptotic,
de2019exploration}. Our simplified theory is however capable of reproducing the trade-off observed in our data (Figure 5f).

\begin{figure}[h!]
\begin{center}
\includegraphics[width=\textwidth]{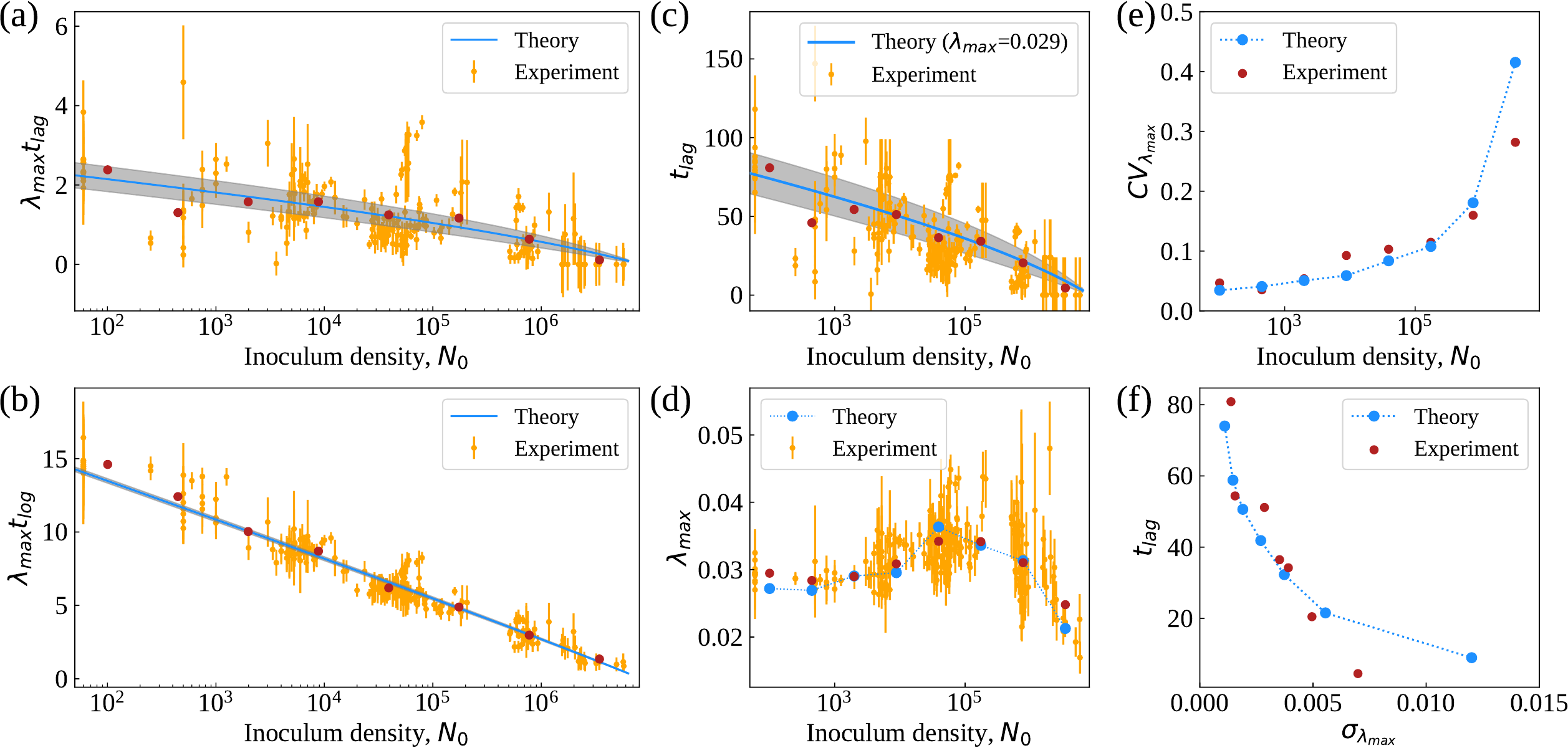}
\end{center}
\caption{(a-b) Behaviour of $\lmax\tlag$ (panel a) and $\lmax\tlog$ (panel b) versus $N_0$ from experiments (orange markers), empirical means (red markers) and theoretical predictions based on Eqs (\ref{ltlag}) and (\ref{ltlog}). Gray shaded areas represent the 95\% confidence intervals. (c-e) Empirical behaviour of $\tlag$ (panel c), $\lmax$ (panel d) and of its relative standard deviation $\CV_{\tlag}$ (panel e) versus $N_0$ in individual experiments (orange markers), together with the empirical means (red markers) and the theoretical predictions based on Eq.~(\ref{ltlag}) with the single adjustable parameter $p$ (blue line in (c) and blue markers in (d) and (e)). See Materials and Methods for details. (f) Empirical relationship between the mean lag time $\tlag$ and the the standard deviation of the maximal growth rate, $\sigma_{\lmax}$ (red markers). Blue markers represent the theoretical prediction based on Eq.~(\ref{ltlag}) (see Materials and Methods for details).}
\end{figure}


\subsection*{Possible role of mechanical interactions and intercellular signaling}

The positive feedback between $\lmax$ and $N_0$ suggests that cooperative effects, e.g. due to mechanical or biochemical interactions, play a role in shaping the growth rate profile.  Notably, Jurkat cells tend to cluster during growth, implying a potential role of mechanical sensing in modulating the fitness. We hence reasoned that, if cooperation is indeed due to mechanical interactions, two populations inoculated at the same density would grow at different rates if cluster formation was inhibited in one population but not in the other. To test this hypothesis, we performed experiments in which clusters were dissolved  by stirring at different frequencies (one or three times per day and every two or five days), see Supplementary Information. The growth rates measured in these conditions were however indistinguishable within the error (see Supplementary Information, Figure S4). This suggests that mechanical interactions do not significantly affect the growth kinetics of Jurkat cells, leaving intercellular signaling e.g. via growth factors as the main suspect. A possible mechanism was suggested in \cite{Archetti2015}, where the proliferation rate of pancreatic cancer cells was shown to be a Hill function of the concentration of insulin-like growth factor II (IGF-II), a hormone excreted by cells during proliferation. Assuming the concentration of growth factors to be proportional to the density of cells at the time at which growth is fastest, which in turn is proportional to $N_0$, one may expect the maximum growth rate to approximately satisfy the relationship 
\begin{equation}\label{lmax2}
\lambda_{\max}\simeq\underbrace{\left[r_0+(\delta r)\frac{N_0^\beta}{N_0^\beta+N_c^\beta}\right]}_{\mathrm{cooperation}}\times\underbrace{\left(1-\frac{N_0}{k}\right)}_{\mathrm{limitation}}~~.
\end{equation}
The factor in square brackets represents the Hill-like cooperative term, with $\beta$ an exponent characterizing the dependence on the initial density, $r_0$ the growth rate in absence of cooperation (i.e. for $N_0$ much smaller than the threshold density $N_c$) and $\delta r$ the maximum extra fitness achievable by cooperation. The term in round brackets is instead the growth-curbing logistic term induced by a finite carrying capacity. Unsurprisingly in view of the many parameters, such an expression provides a very good fit to empirical data (see Supplementary Information, Figure S8). More importantly it suggests that, upon increasing $k$, the average $\lmax$ vs $N_0$ will plateau (as opposed to continuing to increase) before starting to decrease close to the carrying capacity. Such a scenario can be tested experimentally.

\section*{Discussion}

Despite the variability uncovered in recent years in the behaviour of single cells within a population\cite{Wang2010,Kiviet2014}, it is not unreasonable to expect that different populations of the same cells grown in identical media will display similar growth characteristics (e.g. lag times and growth rates) due essentially to the fact that different single-cell features will be averaged out in large enough populations. This however is not necessarily the case\cite{dufrenne1997effect}. In particular, the density of the initial inoculum $N_0$ is known to affect sample-to-sample variability in population growth in unexpected ways\cite{Irwin2010}. To get a large-scale picture of this phenomenon, we have studied how growth rates and lag times of two widely used cancer cell lines grown in a prescribed medium are modulated by $N_0$. Our results detail a previously unobserved scenario. By varying $N_0$ over five orders of magnitude, we found that both the mean lag time and lag time fluctuations tend to decrease monotonically with $N_0$. On the other hand, the maximum growth rate $\lambda_{\max}$ displays a peak at intermediate values of $N_0$, with the largest fluctuations occurring around the peak. While the impact of environmental, genetic or epigenetic factors is not surprising, the complex dependence on $N_0$ that we observe in macroscopic parameters was not expected, both based on logistic growth models and on the Monod-like dependence of the growth rate on other exogenously-controlled parameters \cite{MacIver2008}. 

Given that the growth medium is identical to the one in which cells were pre-grown, the very presence of a lag time points to a conditioning of the pre-growth medium. Both Jurkat and K562 cells are known to produce factors that play pivotal roles for \emph{in vitro} proliferation, like interleukin-2 and prolactin (Jurkat) \cite{Matera1997} and the erythroid-potentiating activity (EPA) glycoprotein (K562) \cite{Avalos1988}. The observed mean lag time is however consistent with a scenario in which seeds in the inoculum adapt independently to the growth medium during the lag phase, albeit with strongly heterogeneous characteristic times. In a limiting case with two sub-populations (fast vs slow adapters), we found that results are reproduced when fast adapters begin to expand shortly after inoculation (see Materials and Methods). 

Understanding the origin of the non-monotonic $N_0$-dependence of $\lambda_{\max}$ is a more subtle issue, especially in view of the expectation that smaller initial densities should lead to larger variability and of the fact that the observed behaviour is not consistent with purely logistic or Allee-like models. We propose that two factors contribute. On one hand, weakly cooperative interactions mediated by excretion of growth factors during growth lead to the positive feedback between $\lmax$ and $N_0$ that is seen at smaller inoculum densities. On the other hand, the finite carrying capacity limits growth starting from higher densities. Based on this, we predict that, if the carrying capacity is augmented, the maximum growth rate versus $N_0$ will achieve a plateau before decreasing (see Supplementary Figure S8). 

Our results can also shed light on a known feature of cancer growth, namely the dependence of the primary and metastatic tumor take rates on the density of cells orthotopically implanted in animal models. In breast cancer \cite{Gregorio2016}, both rates (the latter more markedly so) were found to be higher when the number of implanted cells was smaller, implying more reproducible outcomes for smaller values of $N_0$. Likewise, tumor onset times and doubling times turned out to be negatively correlated with $N_0$, so that smaller inocula implied longer latency periods and slower growth rates. Such a scenario is fully consistent with our results, which strongly suggests that the picture we uncover might hold well beyond the experimental conditions we employed. In this sense, the pattern of variability of the growth rate versus $N_0$ reflects an implicit tradeoff behind the expansion of a population of cancer cells. The longer latency times and slower growth rates deriving from smaller inocula allow for the establishment of more sustained cooperative effects and microenvironmental modifications, possibly driven by the excretion of cytokines and other signaling molecules. This ultimately leads to more efficient take rates. For larger inocula, instead, populations quickly run into limitations in nutrient supply that negatively affect their expansion potential. 

For this reason, it would be particularly important to enrich the set of experimental observations by analyzing these issues in different cell types. Moreover, the identification of the population-level mechanisms underlying this scenario is an open challenge. If the qualitative patterns uncovered in the present work are generic (i.e. not specific to the cell types we investigated), the same mechanisms are likely to be active in different contexts. In this light, a more thorough (and possibly parameter-rich) modeling approach could go  beyond our minimal framework in shedding light on these findings. Developing stochastic models that overcome the conceptual framework of logistic models and account for growth-factor mediated cooperation and finite nutrient availability appears as the most pressing next step along this path.

\section*{Materials and Methods}

\subsection*{Cell cultures}

Jurkat cells (clone E6-1, ATCC) were cultured in complete growth medium (RPMI-1640 Medium (Gibco) supplemented with $10$\% fetal bovine serum) with $1\%$ penicillin/streptomycin in standard growth conditions, i.e. at $37 ^{\circ}$C and in $5$\% CO$_{2}$, following ATCC suggestion ( https://www.lgcstandards-atcc.org/products/all/TIB-152.aspx?geo\_country=fr\#culturemethod ). Cells growing exponentially were then seeded at different concentrations in 6-well plates (Falcon) in a fixed $5$  ml volume of the same medium and counted every $24$ hours until saturation was reached. To assay growth, cells were pipetted gently and $30\,\mu$l of cells and medium were mixed with $30\,\mu$l of methylene blue ($1$\% in PBS). Blue cells corresponded to dead cells while bright round-shaped cells corresponded to live ones. The mixture was then transferred in three single-use Burker's chambers. For each chamber, five phase contrast photos were taken ($10$ to $15$ different photos for each experiment at any given time) with an Axiovert Zeiss inverted microscope (objective 10$\times$). The pictured surfaces corresponded to a volume of $10^{-4}$ ml. The photos were then analyzed via a MATLAB-based image analysis code that implements a built-in function to count live cells (details below). For very low initial seedings (i.e. $N_0 \simeq 60$ cells/ml), $N_0$ was estimated through subsequent dilutions. K562 cells (ATCC) were cultured complete growth medium (RPMI-1640 medium (Gibco) supplemented with $10$\% fetal bovine serum) in standard growth conditions, i.e. at $37 ^{\circ}$C and in $5$\% CO$_{2}$, following ATCC suggestion ( https://www.lgcstandards-atcc.org/products/all/CCL-243.aspx?geo\_country=gb\#culturemethod ). Cells growing exponentially were then seeded at different concentrations in $75$ ml flasks supplied with $10$ ml of the same fresh medium and counted every $24$ hours until saturation level was reached. To assay growth, live cells were monitored through NucleoCounter, a cell counter that integrates imaging and cytometry to count live cells after proper cell staining.

\subsection*{Cell counting algorithm}

The cell counting algorithm we developed takes one micrograph (phase contrast photo) at a time as input. Following a first reduction of the noise, the micrograph is converted into a binary image through thresholding: after the definition of a threshold, value $1$ or white (resp. $0$ or black) is assigned to all pixels with intensities higher (resp. lower) than the threshold. In our case, the threshold was empirically chosen by analysing the intensity of both background and objects of interest, i.e. live cells. After standard image manipulation to optimize object detection and a watershed transform method \cite{Gonzalez2005} to discriminate single cells within clusters, each object formed by at least $10$ pixels organized in a roughly round shape is labeled as a cell. The algorithm recognized circular objects by computing their circularity $c$, defined as $c=P^2/(4 \pi S)$, where $P$ is the perimeter and $S$ the surface area of a given object. For our analysis, we accepted all cases with $0.51 \leq c \leq 1.6$. 
Our counting method is unbiased with respect to the cell density (see Supplementary Information).
Once the total number of cells per image is obtained, the average over $10$ to $15$ repeated measurements per sample is computed and converted into a concentration value. This value represents the time-point of growth curves (see for example Figure 1b).

\subsection*{Growth curve fitting procedure and parameter estimates}

The behaviour of the logarithm of the number $N$ of live cells per ml (normalized by the initial density $N_{0}$) versus time defines the growth curve of the population. To quantitatively determine the different phases of growth, we applied two different methods depending on the value of the initial density $N_0$. 
For $N_0$ smaller than $10^6$ cells/ml (i.e. for initial densities sufficiently smaller than the carrying capacity), we fit growth curves to the sigmoidal function \cite{Zwietering1990} 
\begin{equation}
\ln\frac{N}{N_{0}}=\frac{A}{1+\exp\left[\frac{4\lambda_{\max}}{A}(\tlag-t)+2\right]}\label{eq:fit-log}~~ ,
\end{equation}
with fitting parameters $A$, $t_{\lag}$ and $\lambda_{\max}$. Such parameters represent respectively $\ln(k/N_0)$, lag time and maximum growth rate (see Figure 1a). In practice, $\lmax$ is the slope of the tangent at the inflection point of the growth curve, while $\tlag$ corresponds to the intercept of this tangent with the time axis. Data were fitted by non-linear least squares and the error bars shown in the figures correspond to  standard errors. Results obtained by using other sigmoidal functions were found to be qualitatively identical. As a representative instance, the scenario obtained by fitting a Gompertz function to data is shown in Supplementary Information, Figure S2. 
For the 32 populations with $N_0$ larger than $10^6$ cells/ml (i.e. for initial densities approaching the carrying capacity), lag times become much smaller than the sampling period (24 hours) and fits using sigmoidal functions lose quality. We thus estimated the maximal growth rate $\lambda_{\max}$ as the maximal empirical time derivative between subsequent points along the growth curves ($(\ln(N/N_0)(t+1)-\ln(N/N_0)(t))/dt$). For the carrying capacity $k$, we computed the arithmetic mean between the highest population density achieved by the growth curves and its previous and subsequent values. Finally, we set the lag times $t_{\lag}$ to 0. Errors for $\lambda_{\max}$ and $k$ were estimated by error propagation, while the error on $\tlag$ was manually set to the sampling period. Mean values for the parameter estimates in all figures were evaluated within equally spaced bins over the logarithm of the initial density $N_0$, and were weighted with the inverse estimated variances of the individual parameter estimates within the bins.

\subsection*{$N_0$-dependent maximum growth rate for logistic  and Allee models}

We want to find the behaviour of the maximum growth rate versus $N_0$ for population growth models of the general type 
\beq
\label{eq:sup:allee}
\frac{1}{N}\frac{dN}{dt} = r\left( 1-\frac{N}{k}\right)\left(\frac{N}{k}\right)^b~~,
\eeq
where $r$ is the intrinsic maximal  population growth rate and $b\geq 0$ is a numerical exponent (returning the logistic model for $b=0$ and the weakly cooperative Allee model for $b>0$). To this aim, it suffices to compute the value of $N$ at which the right-hand-side of the above expression  vanishes. Noting that $\frac{d}{dt}\ln(N/N_0)=\frac{1}{N}\frac{dN}{dt}$, this value ($N^\star$) corresponds to the population density at the inflection point of the growth curve, namely $N^\star=kb/(1+b)$. For initial conditions $N_0<N^\star$, the inflection point occurs at some $t>0$ and $\lmax$ is simply obtained by evaluating the $N$-dependent growth rate (right-hand-side of (\ref{eq:sup:allee})) at $N=N^\star$. If $N_0$ is larger than $N^\star$, however, the inflection point occurs at a negative time, implying that the growth rate is maximum at $t=0$. In summary,
\beq
\label{eq:maxallee}
\displaystyle \lambda_{\text{max}} = 
\begin{cases}
r \frac{b^b}{(1+b)^{1+b}} & \text{for }N_0 \leq \frac{b}{1+ b}k \\
r\left(1- \frac{N_0}{k}\right)\left(\frac{N_0}{k}\right)^b & \text{otherwise}
\end{cases}~~
\eeq
for general $b>0$, while  $\lmax=r(1-N_0/k)$ for the purely logistic model ($b=0$). The latter formula corresponds to the blue curve shown in Figure 3c. From a qualitative viewpoint the two models are however very similar since, in both cases, $\lmax$ is largest at small $N_0$ and decreases (monotonically for $b=0$, following a plateau for $b>0$) as $N_0$ increases.

\subsection*{Quantitative relationships between $\tlag$, $\tlog$ and $\lmax$}

We first note that, according to (\ref{eq:fit-log}), the size of the population at time $t=\tlag$ is $N(\tlag)=k^\alpha N_0^{1-\alpha}$, where $\alpha=(1+e^2)^{-1}\simeq 0.12$. Now consider a population of $N_0$ cells (seeds) inoculated in a given volume of the growth medium at time $t=0$ and assume that each cell $i$ undergoes an adaptation phase of duration $\tau_i$ before proliferating. Assuming for simplicity that each seed proliferates independently at rate $\mu$ following adaptation, the overall number of cells in the population at time $t$ reads
\begin{equation}
N(t)=\sum_{i=1}^{N_0}\left[\theta(\tau_i-t)+\theta(t-\tau_i) e^{\mu(t-\tau_i)}\right]~~,
\end{equation}
where $\theta(x)$ is the Heaviside step function ($\theta(x)=1$ for $x>0$, $=0$ for $x<0$, $=1/2$ for $x=0$). We can estimate this quantity at time $\tlag$ for $N_0\gg 1$ under the assumption that a fraction $p$ of cells has adaptation times shorter than $\tlag$ (`fast adapters'). This implies 
\begin{equation}
\frac{N(\tlag)}{N_0}\simeq 1-p+e^{\mu \tlag}\Flag~~~~~,~~~~~
\Flag\equiv\frac{1}{N_0}\sum_{i=1}^{N_0}e^{-\mu\tau_i}\theta(\tlag-\tau_i)~~.
\end{equation}
Note that $\Flag$ depends in principle on both $\mu$ and $N_0$. Imposing consistency with the fitting function, i.e. that the above quantity equals $(k/N_0)^\alpha$, one finds that $\mu$ and $\tlag$ are linked by
\begin{equation}\label{tlag}
\mu \tlag\simeq \ln\left[\left(\frac{k}{N_0}\right)^\alpha-1+p\right]-\ln\Flag~~.
\end{equation}
When $\tau_i\simeq\ovl\tau<\tlag$ for all fast adapters, $-\ln\Flag\simeq\mu\ovl\tau-\ln p$. We can further simplify the picture by assuming that fast adapters expand right from inoculation, i.e. $\ovl{\tau}=0$. Upon identifying $\mu$ with $\lmax$ this immediately yields Eq. (\ref{ltlag}). Working along the same lines one also finds that the growth rate $\mu$ and the time $\tlog$ (exit from the exponential growth phase) are linked by 
\begin{equation}
\label{eq:lmax}
\mu \tlog\simeq\mu \tlag+\ln\frac{k}{N_0}~~,
\end{equation}
which corresponds to Eq. (\ref{ltlog}). These results can then be used to derive estimates for the empirical coefficient of variation of individual macroscopic quantities, specifically by straightforward differentiation (if $x,y>0$ and $y=f(x)$, then $\delta y=f'(x)\delta x$, so $\CV_y\equiv\frac{\delta y}{y}=\frac{x}{y}|f'(x)|\CV_x$). For $\lmax$ and $\tlag$ in particular one gets
\begin{gather}
\label{CVlmax}
\CV_{\lambda_{\max}}= \CV_{\tlog} + \frac{\CV_k}{\lambda_{\max}\tlog} \left[ 1 + \frac{\alpha (k/N_0)^\alpha}{(k/N_0)^\alpha -1 +p}  \right]~~, \\
\label{CVtlag}
\CV_{\tlag}= \CV_{\lambda_{\max}}  + \frac{\CV_k}{\lambda_{\max} \tlag} \frac{\alpha(k/N_0)^\alpha}{(k/N_0)^\alpha -1 +p } ~~.
\end{gather}
To find an optimal value for $p$, we fitted the experimental data for the lag time in Figure 5c (orange markers) to Eq. (\ref{ltlag}) assuming constant $\mu=\lmax=0.029$ (thereby leaving a single adjustable parameter). The best fit (according to minimum weighted least squares method) was obtained for $p=0.38$. This value of $p$ was then used to obtain the theoretical lines shown in Figures 5a and 5b. (For simplicity, in all cases we set $k$ to the value $k = 8.4 \times 10^6$, corresponding to the empirical average carrying capacity.) For the coefficients of variation, we used $p$ and $CV_{t_{\log}}$ (or $CV_{\lambda_{\max}}$) as fitting parameters (see Figures 5e and S9; the agreement improves if $\CV_k$ is also used as a fitting parameter, see Figure S10). Notice that the value of $p$ providing the optimal agreement with experiments is consistent with the empirical behaviour of $-\ln\mathcal{F}_{\lag}$ derived from Eq. (\ref{tlag}). Specifically, the fact that the mean value of $-\ln\mathcal{F}_{\lag}$ is roughly $1$ (Supplementary Figure S11) indeed implies $p\simeq 1/e\simeq 0.3678$ (to be compared with our best estimate of $0.38$.

\section*{Acknowledgements}

The authors would like to thank Carlo Cosimo Campa, Federico Bussolino, Daniele De Martino, Mark Isalan, Roberto Mulet, Lucia Napione and Andrea Pagnani for useful insights. CEB, FC and CB would like to acknowledge the support of the Royal Society International Exchanges 2018 Round 1 grant IES\textbackslash R1 \textbackslash 180027. CEB, ADM and CB acknowledge funding from the Marie Sk{\l}odowska-Curie Action  MSCA-RISE INFERNET (grant agreement \# 734439).

\bibliography{Draft_inoculum_19}

\clearpage


\section*{\Large{Initial cell density encodes proliferative potential in cancer cell populations - Supplementary Information}}

\renewcommand{\thefigure}{S\arabic{figure}}
\setcounter{figure}{0}

\begin{figure}[!h]
\includegraphics[scale=0.8]{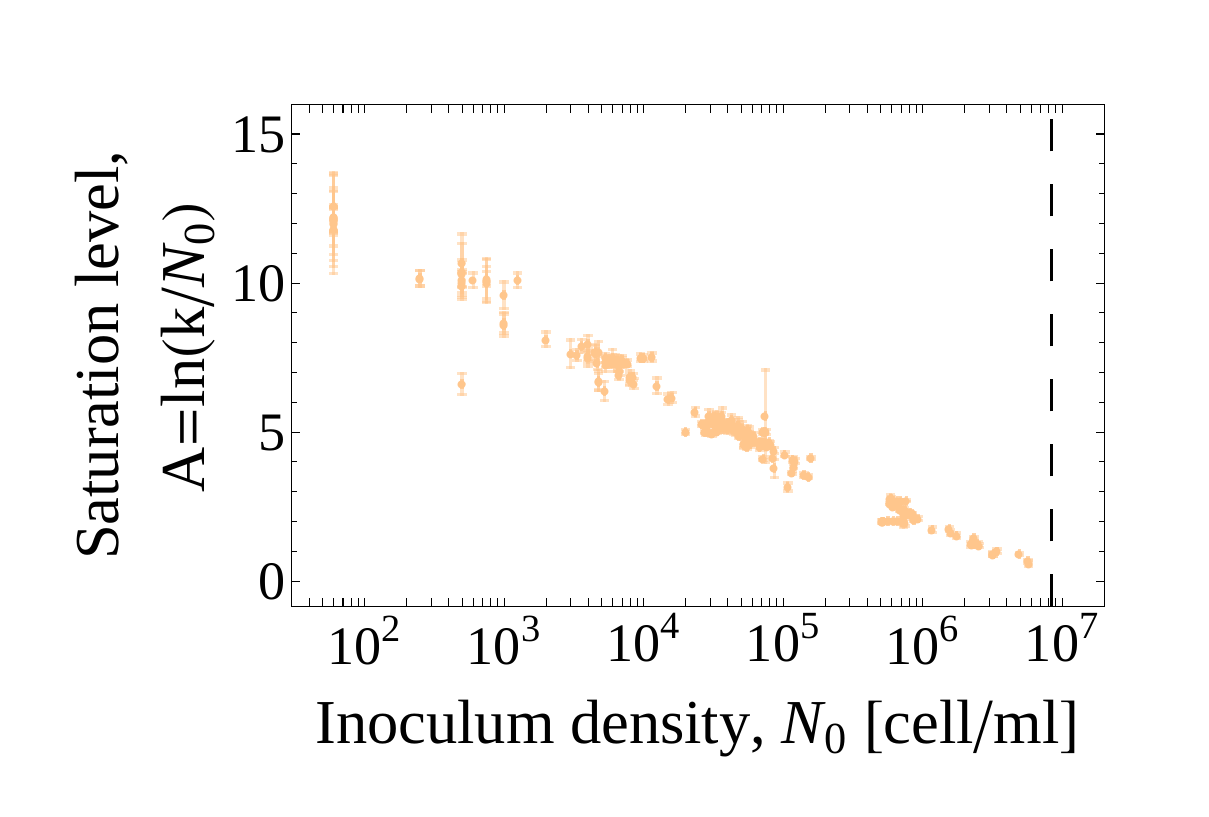}
\caption{The saturation level $A$ of the growth curves (Jurkat cells) is plotted as a function of the inoculum density $N_0$. Each dot corresponds to a specific experiment and carries its own standard error.} 
\end{figure}


\subsection*{Data robustness with respect to the fitting function}

\begin{figure}[!h]
\includegraphics[scale=0.8]{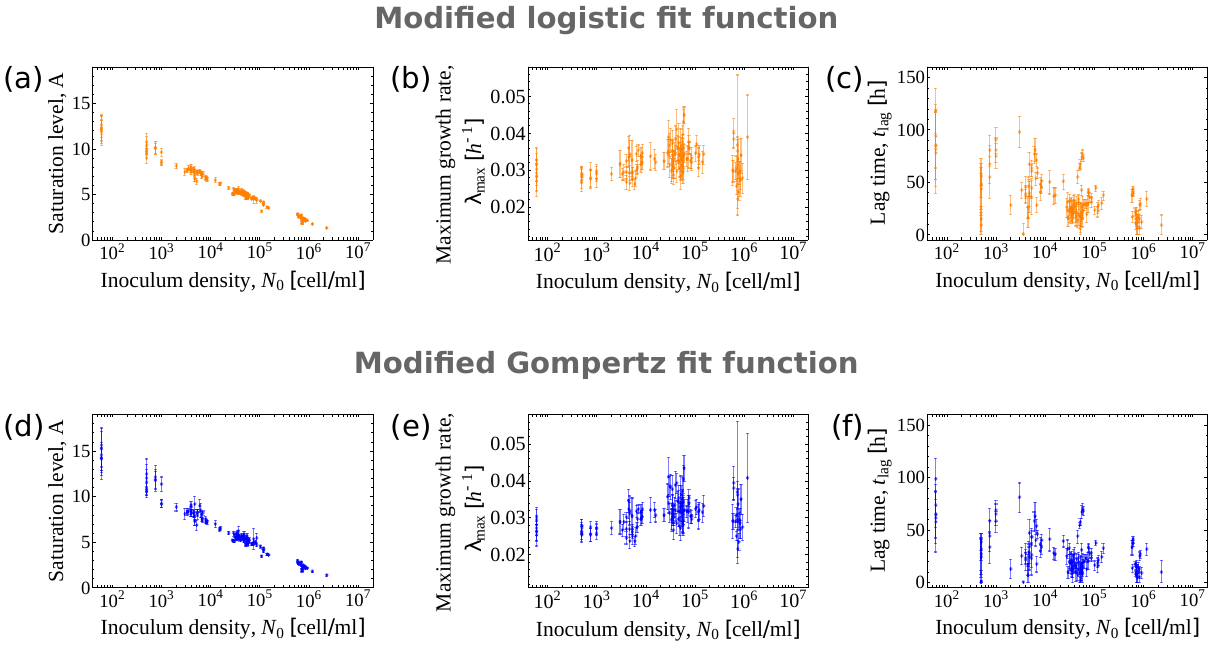}
\caption{Growth parameters obtained by fitting experimental growth curves with the modified logistic function (a-c) and the modified Gompertz function (d-f). The saturation level $A$ (a,d), maximum growth rate $\lambda_{\max}$ (b,e) and lag time $t_{\mathrm{lag}}$ (c,f) are plotted as a function of the inoculum density $N_0$. Each dot corresponds to a specific experiment and carries its own standard error. \label{fig:LogisticGompertz-fit}}
\end{figure}

The modified logistic function used in the main text to fit growth curves did not have any modeling purpose and has been used as a mere tool to quantify the macroscopic variables that characterize growth. To test the robustness of our results, we verified their independence on the fitting function. For such purpose, we fitted the same growth curves with a modified Gompertz function \cite{Zwietering1990} whose parameters are the same of the modified logistic function (maximum growth rate $\lambda_{\max}$, lag time $t_{\mathrm{lag}}$ and growth saturation level $A$):
\begin{equation}
\ln(N/N_{0})= A\exp\left\{-\exp\left[\frac{e\lambda_{\max}}{A}(t_{\mathrm{lag}}-t)+1\right]\right\}\/~~.\label{eq:Gompertz}
\end{equation}
Figure \ref{fig:LogisticGompertz-fit} shows that results obtained by the two fitting procedures are compatible. Our scenario is therefore robust to the use of different fitting functions.

\vspace{2cm}

\clearpage

\subsection*{Comparison between weighted and unweighted growth curve fit}

\begin{figure}[!h]
\includegraphics[scale=0.8]{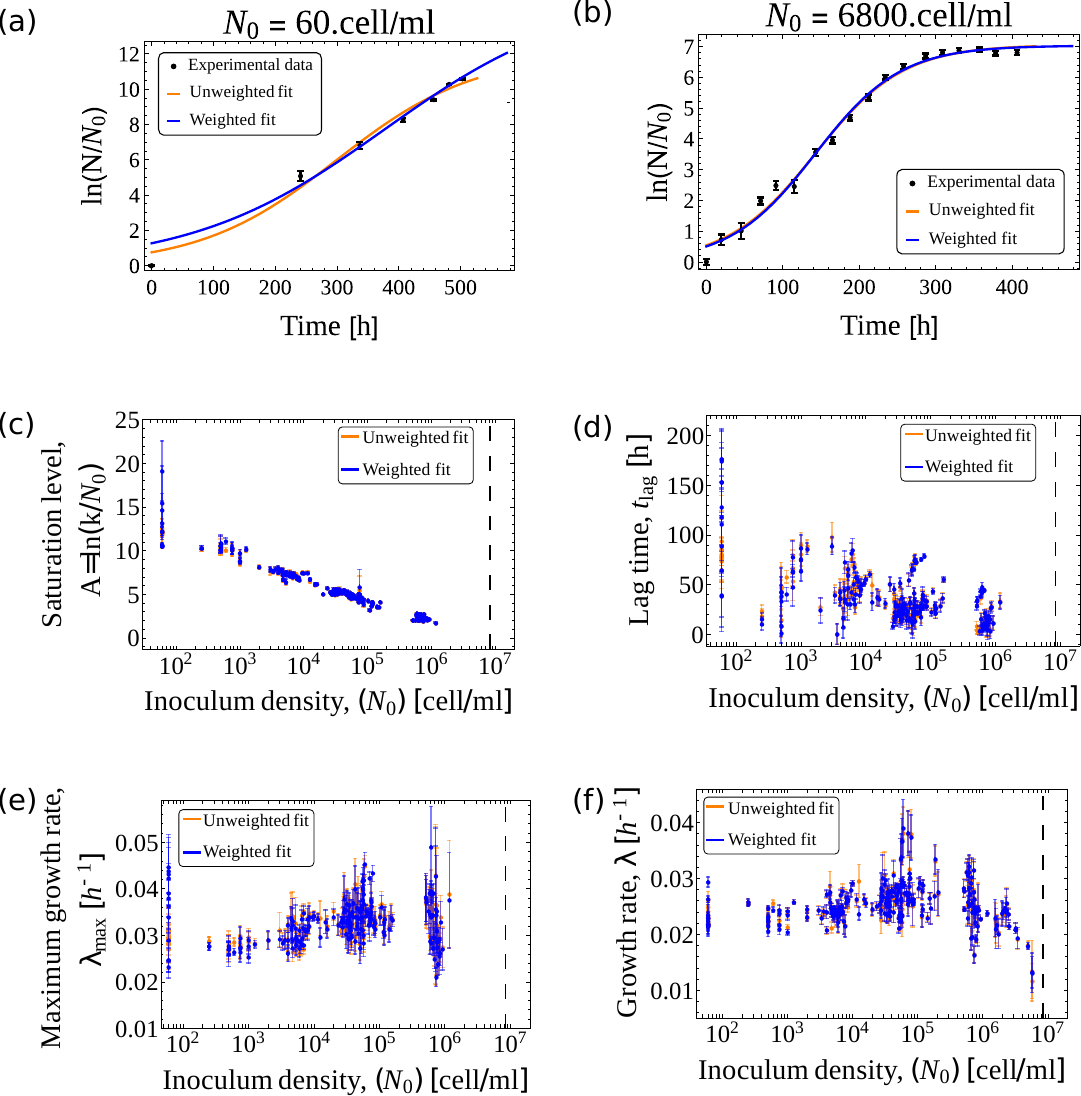}
\caption{Comparison between weighted and unweighted fit of the growth curves. (a-b) Weighted (blue line) and unweighted (red line) fits for two different experiments. Black dots are experimental data together with their standard error of the mean. (c-f) Growth parameters are plotted against $N_0$. The parameters were obtained by performing weighted (blue dots) or unweighted (red dots) fits of the experimental growth curves with the modified logistic function. Each dot corresponds to a specific experiment and carries its own standard error. \label{fig:Weighted-unweighted-fit}}
\end{figure}

All the growth curves that were fitted with a sigmoidal function contained from a minimum of 6 to a maximum of 22 data points. For all these curves we compared parameter estimates from unweighted fits (parameter estimates shown in the main text) to parameter estimates from weighted fits. 
Since each data point within the growth curves is the mean over many subsamples, the weights were measured as the inverse of the variance of the mean of the subsamples at each point (ex. variance of 15 subsample measurements divided by 15) on the log scale.
As it is possible to notice from Figure \ref{fig:Weighted-unweighted-fit}, for most of the growth curves the parameter estimates from weighted and unweighted fits are indistinguishable (normal Z-test showed ~95.5$\%$ of compatibility between the two estimates for all the parameters).

However, parameter estimates from growth curves with low initial density $N_0$ ($N_0<10^2$) looked much more spread for weighted fit than for unweighted fit. This is mainly due to the fact that for these curves a weighted fit does neglect too many points with the risk of doing overfitting. Figure \ref{fig:Weighted-unweighted-fit}a shows exactly this situation: the weighted fit (blue line) clearly disregards the first two points of the curve, thus fitting 5 points with a curve with 3 free parameters. In Figure \ref{fig:Weighted-unweighted-fit}b instead, weighted and unweighted fits look the same. While weighting data points with respect to their error bars makes sense when in presence of many data, in a situation in which all the data are informative, neglecting those that have bigger error bars may be misleading. For this reason we decided to leave in the main text the parameter estimates from unweighted fits.

\clearpage

\subsection*{Stirring experiments}

Stirring experiments, in which clusters of cells were dissolved at regular intervals to test the role of mechanical interactions, were performed with Jurkat cells. Cells were seeded at initial density $N_0\simeq6\cdot10^4$ cells/ml, in the range of value of $N_0$ in which $\lambda_{max}$ increases with the initial density. Three sets of experiments were performed by stirring populations at different  frequencies: (i) three times per day (but counted only once); (ii) once every two days, and (iii) once every five days. In parallel to each of these conditions, a control experiment was run, where cells were stirred and counted once a day. The resulting growth curves are shown in Figure S4b-d, where data represent the mean over experimental replicates and error bars represent their dispersion. The superposition of each experiment (red dots) with its control (blue dots) suggests a negligible influence of clusters in the growth dynamics at the time scales considered.

\vspace{2cm}

\begin{figure}[!h]
\includegraphics[scale=0.8]{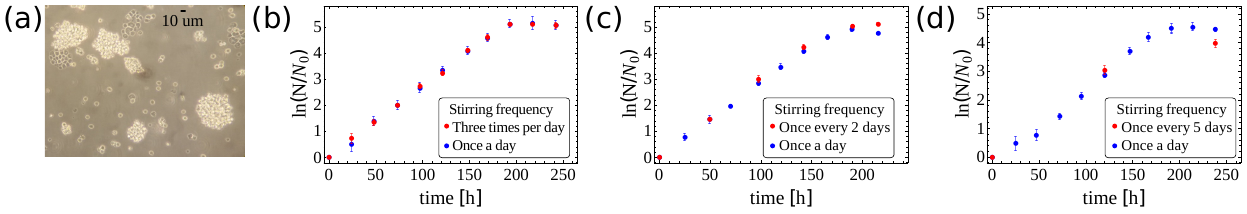}
\caption{Results of stirring experiments. (a) Representative micrograph of Jurkat cell population growing in its standard growth medium. Clusters of cells are visible as well as isolated single cells. During stirring experiments clusters were dissolved at three different  frequencies and counted every day. (b-d) Growth curves obtained from the experiments with stirring frequencies three times a day (b), once every two days (c), and once every five days (d), are shown as red markers. Blue markers represent the control, i.e. no stirring and counting every day. Each marker represents the average over three (b,d) or two (c) replicates, while the error bars denotes the data dispersion over the replicates. \label{fig:Stirring}}
\end{figure}


\newpage{}

\subsection*{Accuracy of the counting algorithm}

\begin{figure}[!h]
\includegraphics[width=\textwidth]{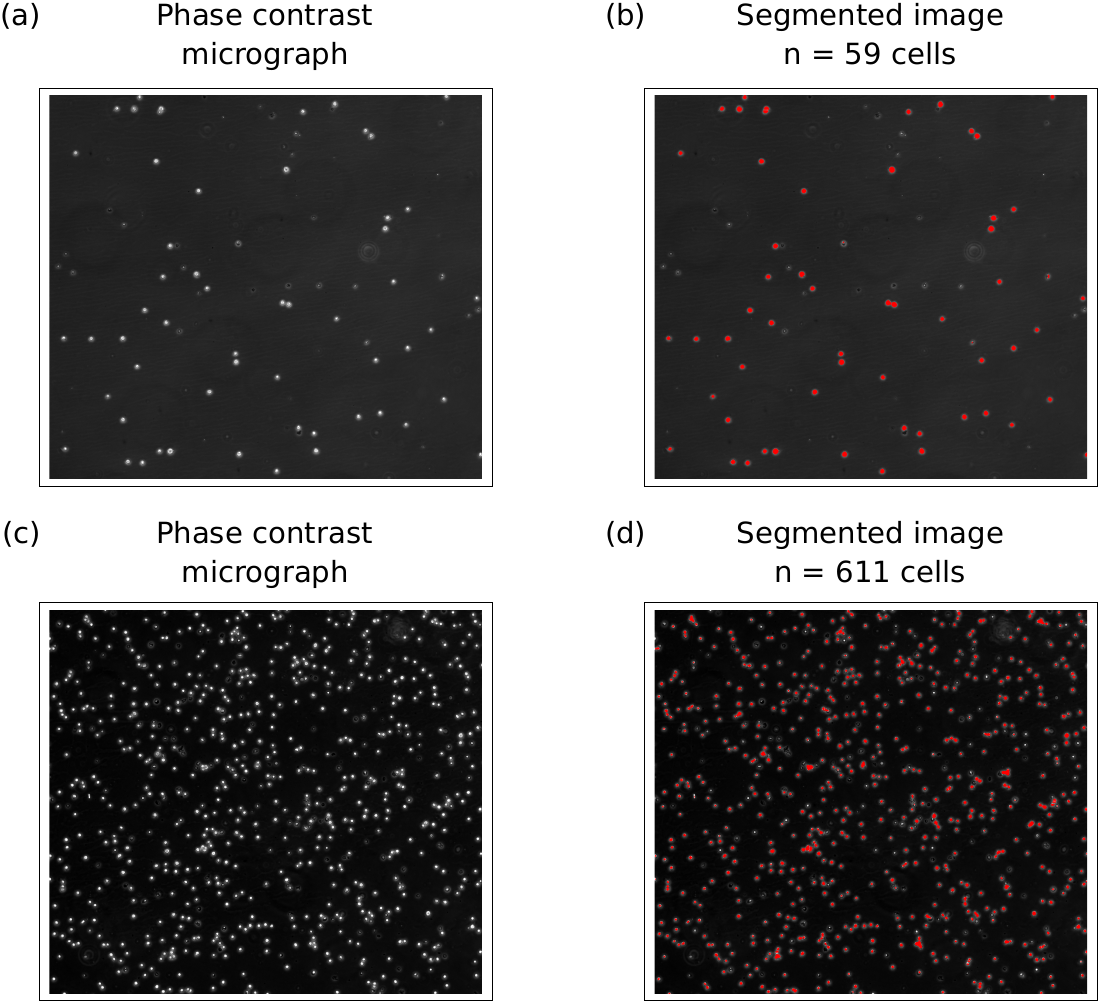}
\caption{Micrographs analysed during the growth experiments. (a,c) Raw phase contrast images showing cells. (b,d) Results of the counting algorithm. Red dots correspond to segmented objects (cells).}
\label{fig:micrograph}
\end{figure}

Figure \ref{fig:micrograph} shows an example of micrographs analysed during the experiments. We showed on the left an example of the raw phase contrast images for two different initial cell densities, while on the right the results of the counting algorithm, where red dots corresponded to the segmented (and counted) objects (cells).

We wondered if the algorithm designed to count cells within micrographs were unbiased with respect to cell density.
To test the accuracy of the algorithm we then automatically generated and analysed images with a known amount of round-shaped objects with the same characteristics of the cells in terms of circularity and number of pixels ($>10$ pixels).

\begin{figure}[!h]
\includegraphics[width=\textwidth]{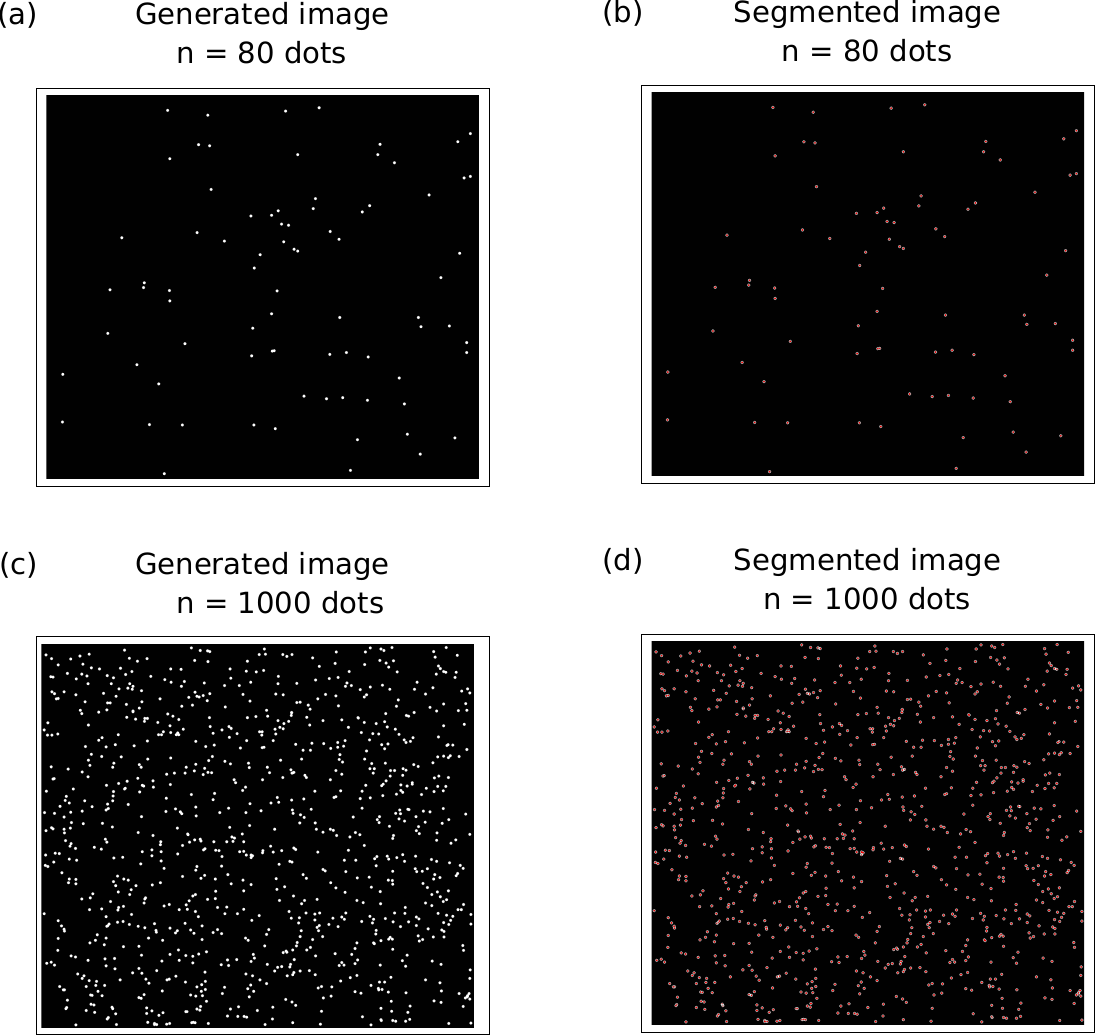}
\caption{(a,c) Images with known amounts of round-shaped objects randomly distributed within the image surface used to test the cell counting algorithm. (b,d) Results of the counting algorithm. Red dots correspond to segmented objects.}
\label{fig:dots}
\end{figure}

The algorithm used to generate images with round-shape objects was custom made and exploited MATLAB built-in functions. Starting from a black image with the same dimensions of the experimental micrographs, the algorithm required as input the radius ($r=4$ pixels) and the number of objects (n) to generate, and then randomly assigned their coordinates before drawing them, see Figure \ref{fig:dots}a,c. 

Figures \ref{fig:dots}b,d shows the result of the segmentation performed by the cell counting algorithm: red areas represent the recognized objects.

\begin{figure}[!h]
\includegraphics[width=\textwidth]{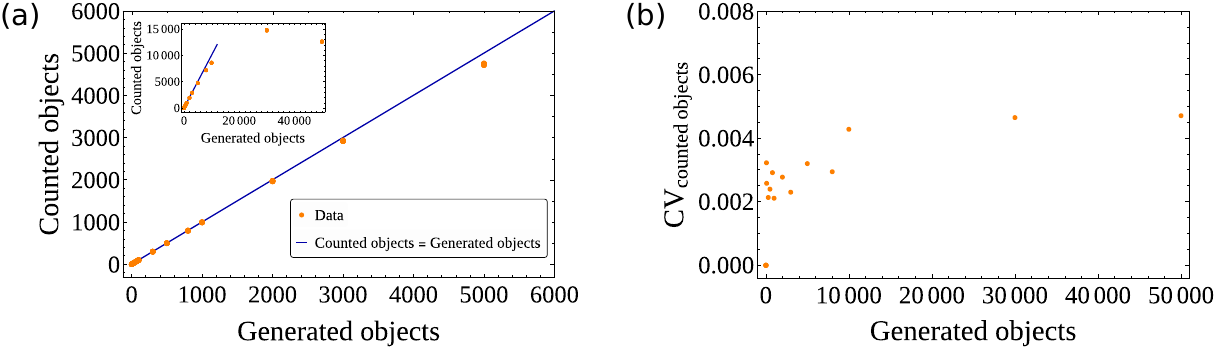}
\caption{(a) Objects counted by the cell counting algorithm are shown against the known amount n of drawn objects. For $n>5000$, data divert from the bisector. (b) CV of the counted objects evaluated over 15 different images as a function of the known amount n of generated objects.}
\label{fig:algorithm}
\end{figure}

To test the accuracy, we then compared the known amount of objects ($n$) generated in the input images (15 images per each n) with the results of the automatic counting ($n'$). 
Figure \ref{fig:algorithm}a shows that for $n \leq 5000$, the values lied on the bisector (blue line in figure), suggesting good counting performance. However, such precision is lost when increasing the number of objects present within the image. This can also be visualized by looking at the CV (ratio between the standard deviation of the counted objects and their mean value): although very small, it increased when increasing n. This is mainly due to the fact that the overpopulation of space obtained by increasing n, led to a higher amount of touching objects (that cannot be resolved by the algorithm). Indeed, to experimentally overcome this issue, before every counting measurement we properly stirred and diluted the cells. Our accuracy test suggested that the algorithm we used to count cells was accurate within the range of cells considered experimentally (the higher amount of cells counted in our experiments was approximately 3000 cells per micrograph).

\begin{figure}[!h]
\includegraphics[width=\textwidth]{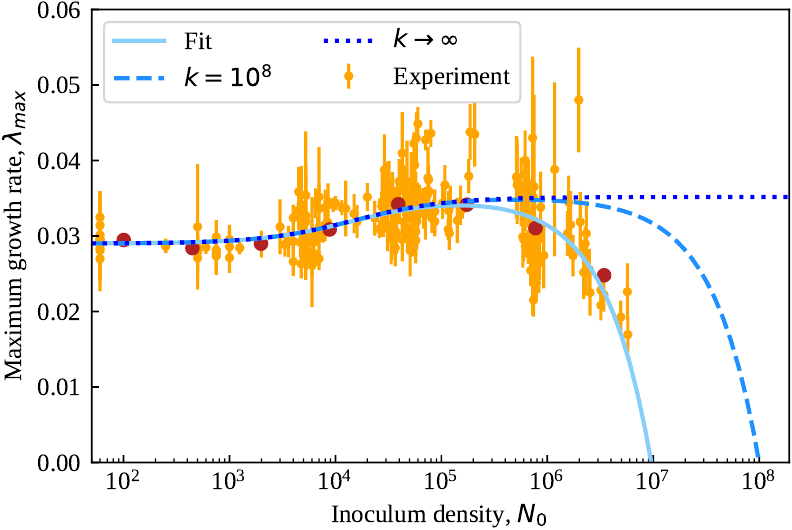}
\caption{Behaviour of $\lambda_{\text{max}}$ versus $N_0$ in individual experiments (orange markers), empirical means (red markers) and best weighted fit of experimental data to Eq.~3 of the Main Text (solid line) with $r_0=0.029$ and $\beta =1$ (constrained), leaving $\delta r$, $N_c$ and $k$ as free parameters. Their optimal values are found to be $\delta r= (6.1 \pm 0.6)\times 10^{-3}$, $N_c= (1.5 \pm 0.5)\times 10^{4}$ and $k= (9.6 \pm 0.9)\times 10^{6}$. The dashed and dotted lines display the behaviour expected on the basis of Eq.~3 of the Main Text for the same value of all parameters except for $k$, which takes values of $k=10^8$ (dashed line) and $k = \infty$ (dotted line), respectively. In essence, an increase of $k$ alone will cause the growth rate to plateau as a function of $N_0$. If cooperation is strengthened, other parameters will also be modified. In such a case, one should expect a further increase of $\lambda_{\max}$ followed by a plateau.}
\end{figure}

\begin{figure}[!h]
\includegraphics[width=\textwidth]{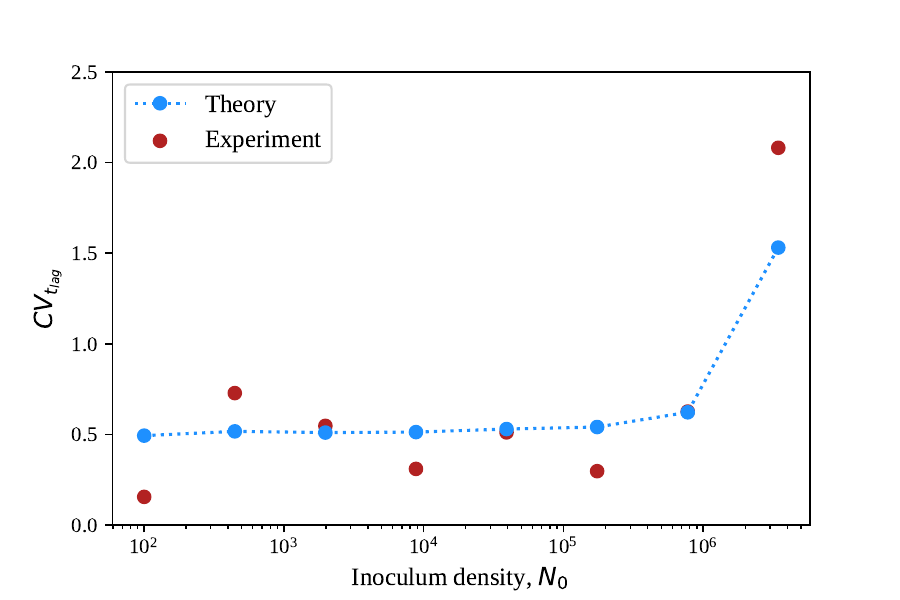}
\caption{Empirical behaviour of 
$CV_{t_{{\rm lag}}}$ versus $N_0$  (red markers) and best fit to Eq. (12) of the Main Text with $p = 0.38$. We used empirical values for $\lambda_{\max}t_{{\rm lag}}$  (varying across bins) and
$CV_k = 0.44$, leaving $CV_{\lambda_{\max}}$  as the only fitting parameter. Its optimal value, $(5 \pm 1) \times 10^{-1}$, is larger than the empirical relative fluctuations of $\lambda_{\max}$. Notice however that in this way we are neglecting the fact that the coefficient of variation of $\lambda_{\max}$ is itself a function of $N_0$. This choice therefore yields a crude approximation for the true $CV_{t_{{\rm lag}}}$.}
\end{figure}


\newpage{}

\begin{figure}[!h]
\includegraphics[width=\textwidth]{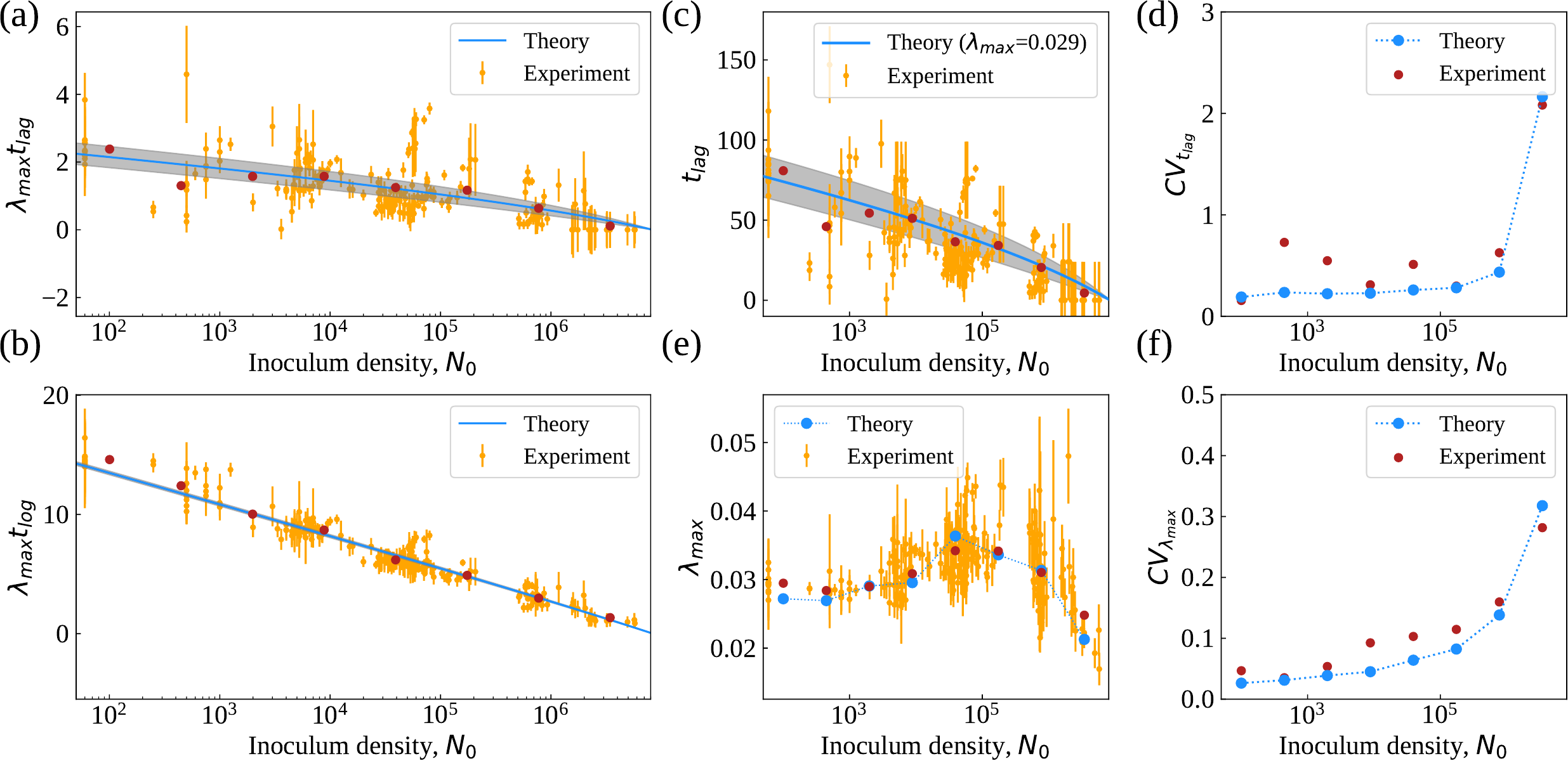}
\caption{(a-b) Behaviour of $\lmax\tlag$ (panel a) and $\lmax\tlog$ (panel b) versus $N_0$ from experiments (orange markers), empirical means (red markers) and theoretical predictions based on Eq. 1 and 2 of the Main Text. For simplicity, we used for $k$ the value $k = 8.4 \times 10^6$, corresponding to the empirical average carrying capacity. (c-d) Empirical behaviour of $\tlag$ (panel c) and $\CV_{\tlag}$ (panel d) versus $N_0$ (orange markers in (c)), empirical means (red markers) and best fit of experimental datapoints to Eq.~1 of the Main Text with $p=0.38$. Grey shaded areas represent the 95\% confidence intervals. For $\CV_{\tlag}$ we used empirical values for $\lmax\tlag$ (varying across bins) and $\CV_{\lambda_{\max}}=0.14$, leaving $\CV_{k}$ as the only  fitting parameter. Its optimal value, $(9\pm 1)\times 10^{-1}$, is of the same order of magnitude of the experimental one. (e-f) Empirical behaviour of $\lmax$ (panel e) and $\CV_{\lmax}$ (panel f) versus $N_0$ (orange markers in (e)), empirical means (red markers) and best fit to Eq.~2 of the Main Text with $p=0.38$ and $\tlog$ set to its empirical value in each bin. For $\CV_{\lmax}$ we used empirical values for $\lmax\tlog$ (varying across bins) and $\CV_{t_{log}}=0$, leaving $\CV_{k}$ as the only  fitting parameter. Its optimal value, $(3.4\pm0.3)\times 10^{-1}$ is of the same order of magnitude as the experimental one.}
\end{figure}


\newpage{}

\begin{figure}[!h]
\includegraphics[width=\textwidth]{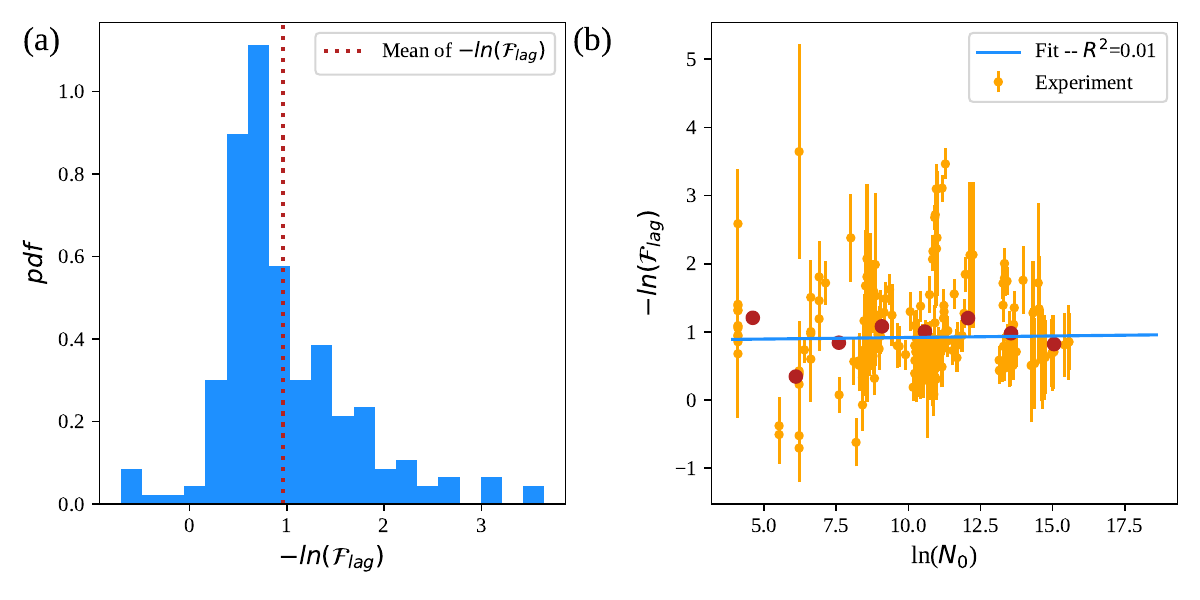}
\caption{(a) Distribution of $-\ln(\mathcal{F}_{\text{lag}})$ computed from experimental data using Eq. (9) of the Main Text with $p=0.38$. The red dotted line marks the empirical mean of the distribution. (b) Experimental data (orange markers), empirical means (red markers) and best linear weighted fit of experimental datapoints (blue line) of $-\ln(\mathcal{F}_{\text{lag}})$ as a function of the logarithm of the inoculum density, $N_0$. The value of $R^2$ suggests that the approximation $-\ln(\mathcal{F}_{\text{lag}})\simeq 1$ employed in our analysis is very accurate.}
\end{figure}


\clearpage

\subsection*{References}

1. Zwietering, M. H., Jongenburger, I., Rombouts, F. M. \& van't Riet, K. Modeling of the bacterial growth curve. \it{Appl. Environ.
Microbiol.} 56, 1875–1881 (1990). URL \red{http://www.ncbi.nlm.nih.gov/pmc/articles/PMC184525/}.



\end{document}